\documentclass[fp]{jpsj3}
\usepackage[dvipdfmx]{graphicx}
\usepackage[dvipdfmx]{color}

\renewcommand{\a}{\alpha}
\renewcommand{\b}{\beta}
\newcommand{\bea}{\begin{eqnarray}}
\newcommand{\eea}{\end{eqnarray}}
\newcommand{\f}[2]{\frac{#1}{#2}}
\newcommand{\eq}{&=&}
\newcommand{\nn}{\nonumber \\ }
\newcommand{\ve}{\varepsilon}
\renewcommand{\d}{\delta}
\newcommand{\area}{\int_{-\infty}^\infty }

\newcommand{\p}{\partial}
\newcommand{\pp}[2]{\f{\p #1}{\p #2}}

\newcommand{\siki}[1]{Eq. (\ref{#1})}

\allowdisplaybreaks[1]

\newcommand{\g}{\gamma}

\allowdisplaybreaks[1]

\title{
Replica Approach for Minimal Investment Risk with Cost 
}

\author{Takashi Shinzato\thanks{shinzato@eng.tamagawa.ac.jp}
}
\inst{
Department of Management Science, College of Engineering, 
Tamagawa University, \\ Machida, Tokyo, 1948610, Japan\\
\smallskip
{\rm (Received February 8, 2018; accepted *** 1, 2018)}
} 

\abst{
{
In the present work, the optimal portfolio minimizing the investment risk with cost is discussed analytically, where this objective function is constructed in terms of two negative aspects of investment, the risk and cost. We note the mathematical similarity between the Hamiltonian in the mean-variance model and {the Hamiltonians in 
the Hopfield model and the Sherrington--Kirkpatrick model} and show that we can analyze this portfolio optimization problem by using replica analysis, 
{and derive} the minimal investment risk with cost and the investment concentration of the optimal portfolio. Furthermore, we validate our proposed method through numerical simulations.
}
}
\kword{mean-variance model, investment risk with cost, investment concentration, Lagrange multiplier method, 
replica analysis}

\begin{document}
\maketitle

\section{Introduction}
{The portfolio optimization problem is one of the most actively researched topics 
in mathematical finance, coming from the theory of diversification investment 
management put forth by Markowitz in his pioneer works in 1952 and 1959 \cite{10.2307/2975974,10.2307/j.ctt1bh4c8h}. 
In mathematical finance, ({especially} operations research),
{investment} optimality in some practical situations has been discussed
\cite{doi:10.1287/mnsc.37.5.519,Rockafellar00optimizationof,Perold:1984:LPO:2909945.2909946},
but only analysis of an {annealed disordered system in} the literature of spin glass has been discussed for the portfolio optimization problem, whereas analysis of the quenching system 
{which is desired} by  rational investors has been given little attention.
Recently, however, 
such analysis of the quenched disordered system {desired} by  rational investors 
in the context of diversified investment 
has started to be investigated using the analytical approaches developed in statistical mechanical informatics
and econophysics \cite{Ciliberti2007,doi:10.1080/14697680701422089,PAFKA2003487,KONDOR20071545,
doi:10.1080/1351847X.2011.601661,10.1371/journal.pone.0134968,10.1371/journal.pone.0133846,1742-5468-2017-2-023301,SHINZATO2018986,
doi:10.7566/JPSJ.86.124804,PhysRevE.94.052307,1742-5468-2016-12-123404,2017arXiv170302777S,PhysRevE.94.062102b,2016arXiv160506840S,doi:10.7566/JPSJ.86.063802,110008689817,1742-5468-2017-12-123402}.
For instance, Ciliberti {\it et al.} examined
the investment risk of the absolute deviation model and 
the expected shortfall model
in the portfolio optimization problem 
with a budget constraint by using replica analysis. 
Specifically, they analyzed 
the typical behavior of the ground state in the limit of absolute zero temperature  (the optimal solution of the portfolio optimization problem)
\cite{Ciliberti2007,doi:10.1080/14697680701422089}.
Pafka {\it et al.} 
compared 
the eigenvalue distribution of the variance-covariance matrix derived from practical data 
with 
the eigenvalue distribution of the 
variance-covariance matrix 
defined by novel variables mapped by Cholesky decomposition and 
discussed three types of investment risks in diversification investment 
\cite{PAFKA2003487}.
Kondor {\it et al.} evaluated 
the relationship between noise and estimated error of each optimal portfolio with respect to several risk models: 
the mean-variance model,
the absolute deviation model, 
the expected shortfall model, and
the max-loss model \cite{KONDOR20071545}.
Caccioli {\it et al.} 
used  replica analysis to determine whether the optimal solution of 
the expected shortfall model with 
ridge regression 
is stable \cite{doi:10.1080/1351847X.2011.601661}.
Furthermore, 
Shinzato {\it et al.}
replaced the portfolio optimization problem including a budget constraint with an inference problem using the Boltzmann distribution 
and derived analytically the trial distribution which can approximate the Boltzmann distribution based on the Kullback--Leibler information criterion {using a belief propagation method}.
They also derived the faster solver algorithm for the optimal solution
using the trial distribution \cite{10.1371/journal.pone.0134968}.
}

{
As described above, various investment models have been examined using replica analysis and {a belief propagation method
in these previous studies\cite{10.1371/journal.pone.0133846,1742-5468-2017-2-023301,SHINZATO2018986,
doi:10.7566/JPSJ.86.124804,PhysRevE.94.052307,1742-5468-2016-12-123404,2017arXiv170302777S,PhysRevE.94.062102b,2016arXiv160506840S,doi:10.7566/JPSJ.86.063802,110008689817,1742-5468-2017-12-123402},
}
but in recent years, attention has been given to the mathematical similarity between 
{the Hopfield model and the most representative investment models, that is mean-variance model}.
For instance, 
Shinzato showed with the Chernoff inequality and replica analysis that the investment risk of the mean-variance model and the investment concentration of the 
{optimal}
 portfolio satisfy the self-averaging property
\cite{10.1371/journal.pone.0133846}.
In addition, Shinzato analyzed the minimization problem of 
investment risk with constraints of budget and 
investment concentration by using replica analysis, 
comparing the results with those of a previous work\cite{10.1371/journal.pone.0133846}, 
as well as analyzing the influence of the investment concentration constraint on the optimal portfolio\cite{1742-5468-2017-2-023301}.
Moreover, Shinzato
further investigated 
the maximization problem of 
investment concentration with constraints of budget and
investment risk in {a previous work\cite{1742-5468-2017-2-023301}} 
and the corresponding minimization problem as a {counterpart},  and
derived the mathematical structures of the two optimal portfolios of 
the primal--dual optimization problems\cite{SHINZATO2018986}.
{Further, Tada {\it et al.} {resolved} the
primal--dual optimization problems by using 
Stieltjes transformation of 
the asymptotical eigenvalue distribution of the Wishart matrix
in order to validate 
the findings {in previous works\cite{1742-5468-2017-2-023301,SHINZATO2018986}} 
where the analysis used replica {analysis\cite{doi:10.7566/JPSJ.86.124804}.}
That is, they 
reexamined 
the minimization problem of investment risk with 
constraints of budget and investment concentration (and the corresponding maximization problem)
and the maximization problem of 
investment concentration with
constraints of budget and investment risk (and the corresponding minimization problem) 
without using replica analysis or the replica symmetry {ansatz.}}
In addition, Shinzato considered the minimization problem of investment risk with constraints of budget and expected return, and the maximization problem of expected return with constraints of budget and investment risk as a primal--dual optimization problem, analyzing 
{them} by using replica analysis and reexamining the relationship between the two optimal portfolios\cite{PhysRevE.94.052307}.
Varga-Haszonits {\it et al.}
{generalized}
the minimization problem of investment risk 
with 
constraints of budget and expected return that was {considered} in {the work by Shinzato}\cite{PhysRevE.94.052307} 
and analyzed the stability 
of the replica symmetry solution\cite{1742-5468-2016-12-123404}.
Shinzato examined the minimization problem of investment risk with constraints of budget and expected return by using replica analysis and derived a macroscopic theory like the Pythagorean theorem of the Sharpe ratio and opportunity loss
\cite{2017arXiv170302777S}.
In addition,
Shinzato analyzed the minimization problem of investment risk with a budget constraint when the variance of the asset return is not unique using replica analysis and a belief propagation method and calculated the minimum investment risk per asset and the investment concentration of the optimal portfolio\cite{PhysRevE.94.062102b}.
Furthermore,
{using the asymptotic eigenvalue distribution of the Wishart matrix,} Shinzato {in a previous work\cite{2016arXiv160506840S}} 
reexamined 
the minimization problem of investment risk per asset with 
constraints of budget and investment concentration {of the optimal portfolio handled} in 
the earlier work \cite{PhysRevE.94.062102b}.
As a related case,
Shinzato examined the 
minimization problem of 
investment risk with a budget constraint
when 
the return is characterized by a single-factor model
by using replica analysis and 
succeeded in quantifying the influence of common factors included in the minimal investment risk
\cite{doi:10.7566/JPSJ.86.063802}.
Moreover, 
Shinzato examined the minimization problem of 
investment risk with constraints of budget and 
short-selling by using replica analysis when the asset 
returns are independently and identically distributed, and 
confirmed that the minimal investment risk per asset based 
on the replica symmetric ansatz {has} a first-order phase transition\cite{110008689817}.
{Following Shinzato's results, Kondor {\it et al.} examined} the problem of the minimization {of} 
a specific type of risk function with constraints of budget and short-selling  
by using replica analysis for the case that {each} asset return is not necessarily distributed identically for all assets, and clarified that their minimal risk function has a first-order phase transition
\cite{1742-5468-2017-12-123402}.
}

{
As described above, at various investment opportunities, objective criteria (such as investment risk, purchase cost, expected return, and investment concentration) that rational investors hope to know have been examined using the approaches of a quenched disordered system (e.g., replica analysis and a belief propagation method).
However, it is also known that rational investors do not directly use only these objective criteria, but rather investment activities are carried out based on each investor's utility function \cite{Luenberger1998InvestmentScience,marcus2014investments}.
Such a utility function is based on investment preferences ({namely,  
risk averse/risk neutral/risk loving}) of each investor, and furthermore, the utility function involves a combination of investment risk, purchase cost, and expected return.
Among the previous cross-disciplinary research, few studies discussed  the utility function, so it has been difficult to build a theory that appropriately supports investment decisions by rational investors.
}

{Therefore,
in order to provide a seamless connection between 
the analytical approach discussed in previous {works\cite{Ciliberti2007,doi:10.1080/14697680701422089,PAFKA2003487,KONDOR20071545,
doi:10.1080/1351847X.2011.601661,10.1371/journal.pone.0134968,10.1371/journal.pone.0133846,1742-5468-2017-2-023301,SHINZATO2018986,
doi:10.7566/JPSJ.86.124804,PhysRevE.94.052307,1742-5468-2016-12-123404,2017arXiv170302777S,PhysRevE.94.062102b,2016arXiv160506840S,doi:10.7566/JPSJ.86.063802,110008689817,1742-5468-2017-12-123402}}
and the analysis of utility functions, that is,
as a first step of an analysis of utility functions,
{we} examine the minimization problem of 
a loss function defined by two objective criteria under a budget constraint
by using replica analysis.
In particular,
we assume 
the utility function of the
rational investors whose hope is to reduce 
two negative aspects of 
investment, the investment risk (fluctuation risk of the held asset occurring during the investment period) and purchasing (or selling) cost (cost incurred in investing).
}

{
The remainder of the paper is organized as follow.  
In the next section, 
the portfolio optimization problem with a budget constraint 
for minimizing the loss function defined by the investment risk and purchasing cost (which we refer to {hereafter} as the investment risk with cost) is formulated.
Section \ref{sec3}
demonstrates that 
the computation complexity for 
finding the optimal portfolio minimizing the 
investment risk with cost 
is increasing with the number of assets 
by an analysis of this portfolio optimization problem with the Lagrange multiplier method,
and therefore that it is difficult to evaluate this problem in practical situations.
In section \ref{sec4}, 
with the aim of avoiding this computational difficulty when using the Lagrange multiplier method,
we assess the minimal investment risk with cost per asset
and its investment concentration 
by using replica analysis. 
Further, we 
compare the findings obtained by our proposed method with the minimal expected investment risk with cost 
and its investment concentration derived from the analytical procedure in previous work.
In section \ref{sec5},
the effectiveness of our proposed method is 
verified by numerical simulations. 
The final section is devoted to summarizing the present study and discussing future research.
}

\section{Model Setting\label{sec2}}
{
In the present work, 
we consider 
the portfolio optimization 
problem 
with a budget constraint 
in which one invests in $N$ assets at each of $p$ periods 
in a stable investment market with no restrictions on short-selling
and show the properties of the optimal portfolio minimizing the 
objective function defined by the two loss functions capturing the negative aspects of investment, investment risk and purchasing cost. 
First, 
the portfolio of asset $i(=1,2,\cdots,N)$ is $w_i\in{\bf R}$, 
and the portfolio of all $N$ assets is $\vec{w}=(w_1,w_2,\cdots,w_N)^{\rm T}\in{\bf R}^N$.
The notation ${\rm T}$ indicates the transpose of a vector or matrix and, using the same setting as in previous {works
\cite{Ciliberti2007,doi:10.1080/14697680701422089,PAFKA2003487,KONDOR20071545,
doi:10.1080/1351847X.2011.601661,10.1371/journal.pone.0134968,10.1371/journal.pone.0133846,1742-5468-2017-2-023301,SHINZATO2018986,
doi:10.7566/JPSJ.86.124804,PhysRevE.94.052307,1742-5468-2016-12-123404,2017arXiv170302777S,PhysRevE.94.062102b,2016arXiv160506840S,doi:10.7566/JPSJ.86.063802,110008689817,1742-5468-2017-12-123402}} 
, the budget constraint of the portfolio $\vec{w}$ is defined as
\bea
\label{eq1}
\sum_{i=1}^Nw_i\eq N.
\eea
In addition, 
the return of asset 
$i$ at period $\mu(=1,2,\cdots,p)$ is represented by $\bar{x}_{i\mu}$, 
and is independently distributed according to some distribution with mean $E[\bar{x}_{i\mu}]=r_i$ and variance $V[\bar{x}_{i\mu}]=v_i$. Moreover, 
purchasing cost per portfolio of asset $i$ at 
the first period of investment is $c_i$. Using this notation, 
the investment risk and total purchasing cost are giving by
\bea
\label{eq2}
Risk
\eq
\f{1}{2N}\sum_{\mu=1}^p
\left(
\sum_{i=1}^Nw_i\bar{x}_{i\mu}-
\sum_{i=1}^Nw_ir_i
\right)^2\nn
\eq\f{1}{2}\sum_{i=1}^N\sum_{j=1}^Nw_iw_j
\left(
\f{1}{N}\sum_{\mu=1}^px_{i\mu}x_{j\mu}
\right)
,\\
\label{eq3}Cost\eq\sum_{i=1}^Nw_ic_i.
\eea
Since the first term of the first line in \siki{eq2}, $\sum_{i=1}^Nw_i\bar{x}_{i\mu}$, describes 
the {total return}  at period $\mu$
and the second term $\sum_{i=1}^Nw_ir_i$ represents its expectation, the investment risk is defined by 
the sum of 
the squared of differences between 
the {total return} at each period, $\sum_{i=1}^Nw_i\bar{x}_{i\mu}$, and 
the expected {total return} $\sum_{i=1}^Nw_ir_i$. Further, for the sake of simplicity,
here the modified return $x_{i\mu}=\bar{x}_{i\mu}-r_i$ is {used}; 
note that the mean 
and the variance of the modified return $x_{i\mu}$ are $E[x_{i\mu}]=0$ and $V[x_{i\mu}]=v_i$, respectively.
\siki{eq3}
represents the total purchasing cost.
}

{
Based on the above {model setting}, as the objective function, 
using the cost tolerance $\eta(>0)$, the investment risk plus the total purchasing cost at the first period of investment 
is represented as ${\cal H}(\vec{w}|X,\vec{c})=Risk+\eta\times Cost$ (and is what we are calling the investment risk with cost), and is expressed as
\bea
\label{eq4}
{\cal H}(\vec{w}|X,\vec{c})
\eq\f{1}{2}\vec{w}^{\rm T}J\vec{w}+\eta\vec{c}^{\rm T}\vec{w},
\eea
where 
the variance-covariance matrix (that is, the Wishart matrix) defined by the modified return $x_{i\mu}$,
$J=\left\{J_{ij}\right\}\in{\bf R}^{N\times N}$,
and cost vector $\vec{c}=(c_1,c_2,\cdots,c_N)^{\rm T}\in{\bf R}^N$
are 
used in \siki{eq4}. Specifically, 
the $(i,j)$th component of Wishart matrix $J$ is $J_{ij}=\f{1}{N}\sum_{\mu=1}^px_{i\mu}x_{j\mu}$.
Moreover, using return matrix 
$X=\left\{\f{x_{i\mu}}{\sqrt{N}}\right\}\in{\bf R}^{N\times p}$,
$J=XX^{\rm T}$ is also defined.
From the definition of the investment risk with cost in \siki{eq4},
cost tolerance $\eta$  
is the tolerance degree  of the investor with respect to the added cost.
}

{
One point should be noticed here. 
The investment risk with cost discussed in this work, ${\cal H}(\vec{w}|X,\vec{c})$, is 
regarded as the Hamiltonian in this investment system, which allows us
to apply 
several analytical approaches developed 
in spin glass theory 
to analyze
the typical behaviors of the optimal portfolio of this portfolio optimization problem multidirectionally.
The reason for this is that, given 
$N$ Ising spins $\vec{S}=(S_1,S_2,\cdots,S_N)^{\rm T}\in\left\{\pm 1\right\}^N$
and extremal magnetic field 
$\vec{h}=(h_1,h_2,\cdots,h_N)^{\rm T}\in{\bf R}^N$,
square symmetric matrix $J$
represents Hebb's law  in the case of the Hopfield model {and/or} the RKKY interaction matrix in the case of the Sherrington-Kirkpatrick (SK) model.
The Hamiltonian of the Hopfield or SK model
${\cal H}(\vec{S})$ is defined by
\bea
\label{eq5}
{\cal H}(\vec{S})
\eq
-\sum_{i>j}J_{ij}S_iS_j-\sum_{i=1}^Nh_iS_i\nn
\eq-\f{1}{2}\vec{S}^{\rm T}J\vec{S}-\vec{h}^{\rm T}\vec{S},
\eea
{where the notation $\sum_{i>j}$ means the sum over all pairs $(i,j)$ satisfying $i>j$.}
Comparing 
Eqs. (\ref{eq4})
 and (\ref{eq5}), 
it is easily seen that {they  are mathematically similar with respect to these two models.} Moreover, 
Wishart matrix $J=XX^{\rm T}\in{\rm R}^{N\times N}$ defined in \siki{eq4}
is related to Hebb's law in the Hopfield model and
the aim of both problems is to minimize the Hamiltonian. Therefore, 
using replica analysis and belief propagation developed in 
fields engaged in cross-disciplinary research  such as spin glass theory and statistical mechanical informatics, 
we can analyze the portfolio optimization problem and 
derive several novel 
insights for 
diversification investment theory.
{That is, 
the optimal portfolio minimizing the 
Hamiltonian constructed from RKKY interaction terms only (i.e., $\eta=0$)
by using the analytical approach for a quenched disordered system
has been investigated in 
previous works, where it has been shown that it is difficult to 
analyze the quenched disordered system 
(i.e., the rational investors can be regarded as in spin glass theory)
by using the analytical approach developed in operations research (i.e., the approach for an annealed disordered system).
\cite{Ciliberti2007,doi:10.1080/14697680701422089,PAFKA2003487,KONDOR20071545,
doi:10.1080/1351847X.2011.601661,10.1371/journal.pone.0134968,10.1371/journal.pone.0133846,1742-5468-2017-2-023301,SHINZATO2018986,
doi:10.7566/JPSJ.86.124804,PhysRevE.94.052307,1742-5468-2016-12-123404,2017arXiv170302777S,PhysRevE.94.062102b,2016arXiv160506840S,doi:10.7566/JPSJ.86.063802,110008689817,1742-5468-2017-12-123402}.
As the natural extension of 
previous works\cite{10.1371/journal.pone.0133846,1742-5468-2017-2-023301,SHINZATO2018986}, 
}we here add terms of external magnetic fields to {investment risk}, that is,  
the total cost, 
in order to {attempt to construct and}
analyze a  utility function
and thereby create a 
macroscopic theory, which would enrich the theory of optimal investment risk.}

{
Under the above assumptions, in the limit of a large number of assets $N$,
the minimal investment risk with cost per asset $\ve$
is 
\bea
\label{eq6}
\ve\eq\lim_{N\to\infty}
\f{1}{N}
\mathop{\min}_{\vec{w}\in{\cal W}}
{\cal H}(\vec{w}|X,\vec{c}),
\eea
where 
the feasible subset of portfolio $\vec{w}$,  
${\cal W}=\left\{\vec{w}\in{\bf R}^N\left|
\vec{w}^{\rm T}\vec{e}=N
\right.\right\}$ and 
the vector of ones
$\vec{e}=(1,1,\cdots,1)^{\rm T}\in{\bf R}^N$
are used. 
{From a previous work\cite{10.1371/journal.pone.0133846}, 
this minimal investment risk with cost 
satisfies the property }
of self-averaging.
Moreover, 
from the definition of 
\siki{eq6},
{the minimal investment risk with} cost is related to the analysis of a quenched disordered system.
On the other hand, from the literature of operations research,
the
minimal expected investment risk with cost per asset 
$\ve^{\rm OR}$ is 
\bea
\label{eq7-1}
\ve^{\rm OR}\eq\lim_{N\to\infty}
\f{1}{N}
\mathop{\min}_{\vec{w}\in{\cal W}}
E_X[{\cal H}(\vec{w}|X,\vec{c})],
\eea
where $E_X[g(X)]$
is the configuration average of the function $g(X)$.
{Equation
(\ref{eq7-1})} shows that this description is related to
the analysis of an annealed disordered system.
Therefore, 
the goal of the present work 
is also to derive and {examine} the 
optimal investment strategy of the portfolio optimization problem 
with rational investors,
so we will discuss $\ve$ in \siki{eq6} in detail, but not $\ve^{\rm OR}$ in \siki{eq7-1}.
}

\section{Lagrange Multiplier Method\label{sec3}}
{
Here, given return matrix $X=\left\{\f{x_{i\mu}}{\sqrt{N}}\right\}\in{\bf R}^{N\times p}$
and using the Lagrange multiplier method,
the minimal investment risk with cost per asset $\ve$ and 
its investment concentration $q_w$ are analytically evaluated.
Lagrange function ${\cal L}(\vec{w},k)$ 
for the minimization problem of the investment risk with cost in 
\siki{eq4}, ${\cal H}(\vec{w}|X,\vec{c})$ under the budget constraint in \siki{eq1},
is defined by
\bea
\label{eq8-1}
{\cal L}(\vec{w},k)\eq
\f{1}{2}
\vec{w}^{\rm T}J\vec{w}+\eta\vec{c}^{\rm T}\vec{w}+
k(N-\vec{w}^{\rm T}\vec{e}),
\eea
where auxiliary variable $k$ is the 
Lagrange multiplier variable with respect to 
the budget constraint in \siki{eq1}.
}

{
The extremum of 
${\cal L}(\vec{w},k)$ satisfies 
$\pp{{\cal L}(\vec{w},k)}{\vec{w}}=0$ and $\pp{{\cal L}(\vec{w},k)}{k}=0$,
so the minimal investment risk with cost per asset is 
\bea
\ve\eq\f{1}{2}
\f{
\left(1+\f{\eta}{N}\vec{e}^{\rm T}J^{-1}\vec{c}
\right)^2
}{
\f{1}{N}\vec{e}^{\rm T}J^{-1}\vec{e}
}
-\f{\eta^2}{2
}\f{1}{N}\vec{c}^{\rm T}J^{-1}\vec{c}.
\label{eq8}
\label{eq9-1}
\eea
Moreover, the investment concentration $q_w=\f{1}{N}\sum_{i=1}^N(w_i^*)^2$ of 
the optimal portfolio 
$\vec{w}^*=\arg\mathop{\min}_{\vec{w}\in{\cal W}}{\cal H}(\vec{w}|X,\vec{c})=(w_1^*,w_2^*,\cdots,w_N^*)^{\rm T}\in{\bf R}^N$
is \bea
q_w\eq\f{\vec{e}^{\rm T}J^{-2}\vec{e}}{N}
\left(
\f{N}{\vec{e}^{\rm T}J^{-1}\vec{e}}
+\eta\left(
\f{\vec{e}^{\rm T}J^{-1}\vec{c}}
{\vec{e}^{\rm T}J^{-1}\vec{e}}
-\f{\vec{e}^{\rm T}J^{-2}\vec{c}}
{\vec{e}^{\rm T}J^{-2}\vec{e}}
\right)
\right)^2\nn
&&+\eta^2
\f{\vec{e}^{\rm T}J^{-2}\vec{e}}{N}
\left(
\f{\vec{c}^{\rm T}J^{-2}\vec{c}}
{\vec{e}^{\rm T}J^{-2}\vec{e}}
-\left(\f{\vec{e}^{\rm T}J^{-2}\vec{c}}
{\vec{e}^{\rm T}J^{-2}\vec{e}}\right)^2
\right).
\label{eq9}
\label{eq10-1}
\eea
In the evaluation of 
the minimal investment risk with cost per asset $\ve$
and 
the investment concentration of the optimal portfolio $q_w$, 
we
need to 
assess 
six moments, 
$\f{1}{N}\vec{e}^{\rm T}J^{-1}\vec{e}$,
$\f{1}{N}\vec{e}^{\rm T}J^{-1}\vec{c}$,
$\f{1}{N}\vec{c}^{\rm T}J^{-1}\vec{c}$,
and 
$\f{1}{N}\vec{e}^{\rm T}J^{-2}\vec{e}$,
$\f{1}{N}\vec{e}^{\rm T}J^{-2}\vec{c}$,
$\f{1}{N}\vec{c}^{\rm T}J^{-2}\vec{c}$, and {also} the inverse matrices $J^{-1}$ and $J^{-2}$.
However, computing these inverse matrices accurately requires an $O(N^3)$ computation.
Thus, we have the problem that 
as the number of assets $N$
becomes larger, {of course, so does the
computation complexity.}
As the number of assets is {typically} $N=10^3$ to $10^5$, 
it is not easy to assess directly either
$\ve$ in \siki{eq9-1}
or $q_w$ in \siki{eq10-1}.
In the following section, therefore, we avoid 
the computation of the inverse of the Wishart matrix
{and propose a} method for effectively analyzing 
the minimal investment risk with cost and 
the investment concentration of the optimal solution.
}

\section{Replica Analysis\label{sec4}}
{Here, following previous {works\cite{Ciliberti2007,doi:10.1080/14697680701422089,PAFKA2003487,KONDOR20071545,
doi:10.1080/1351847X.2011.601661,10.1371/journal.pone.0134968,10.1371/journal.pone.0133846,1742-5468-2017-2-023301,SHINZATO2018986,
doi:10.7566/JPSJ.86.124804,PhysRevE.94.052307,1742-5468-2016-12-123404,2017arXiv170302777S,PhysRevE.94.062102b,
2016arXiv160506840S,doi:10.7566/JPSJ.86.063802,110008689817,1742-5468-2017-12-123402},
}we consider 
the minimal investment risk with cost per asset $\ve$
and its investment concentration $q_w$
in terms of replica analysis. First,
${\cal H}(\vec{w}|X,\vec{c})$ in \siki{eq4} 
is regarded as the Hamiltonian of this 
investment system. 
The partition function of 
the investment market (at inverse temperature $\b$), 
$Z(X)$, is defined by
\bea
Z(X)\eq\int_{{\cal W}}
d\vec{w}e^{-\b{\cal H}(\vec{w}|X,\vec{c})},
\eea
where 
${\cal W}$ is 
the subspace of feasible 
portfolios in \siki{eq1}.
Furthermore,
using this description of the partition, 
from the identity function
\bea
\label{eq11}
\ve\eq-\lim_{\b\to\infty}\left\{
\pp{}{\b}
\lim_{N\to\infty}\f{1}{N}
E_X[\log Z(X)]
\right\},
\eea
{it is known that}
the typical behavior of the minimal investment risk with
cost per asset can be {evaluated\cite{1742-5468-2017-12-123402}.}
Similar to in {this} previous work, 
in order to assess the
configuration average of the logarithm of the partition function 
$E_X[\log Z(X)]$, we 
need to analyze 
the $n$th moment $E_X[Z^n(X)]$ at $n\in{\bf Z}$. That is,
\bea
\label{eq13}
&&\lim_{N\to\infty}
\f{1}{N}\log E_X[Z^n(X)]\nn
\eq
\mathop{\rm Extr}_{\Theta}
\left\{
\f{1}{2}{\rm Tr}Q_w\tilde{Q}_w
+\f{1}{2}{\rm Tr}Q_s\tilde{Q}_s
-\vec{k}^{\rm T}\vec{e}\right.\nn
&&
-\f{\a}{2}\log\det\left|I+\b Q_s\right|
-\f{1}{2}\left\langle
\log\det\left|
\tilde{Q}_w+v\tilde{Q}_s
\right|
\right\rangle\nn
&&
\left.
+
\f{1}{2}\left\langle
\left(\vec{k}-\b\eta c\vec{e}\right)^{\rm T}
\left(
\tilde{Q}_w+v\tilde{Q}_s
\right)
\left(\vec{k}-\b\eta c\vec{e}\right)
\right\rangle
\right\}
,\nn
\eea
is expanded, where 
$Q_w=\left\{q_{wab}\right\}$
and $Q_s=\left\{q_{sab}\right\}
$ are order parameters (with auxiliary parameters $
\tilde{Q}_w=\left\{\tilde{q}_{wab}\right\},\tilde{Q}_s=\left\{\tilde{q}_{sab}\right\}\in{\bf R}^{n\times n},\vec{k}=(k_1,k_2,\cdots,k_n)^{\rm T}\in{\bf R}^n,(a,b=1,2,\cdots,n)
$). Then
the set of order parameters is 
$\Theta=\left\{Q_w,Q_s,\tilde{Q}_w,\tilde{Q}_s,\vec{k}\right\}$.
Moreover, 
the notation 
${\rm Extr}_{m}g(m)$
means 
the extremum of $g(m)$ with respect to $m$, and 
the period ratio is 
$\a=p/N\sim O(1)$ and $\vec{e}=(1,1,\cdots,1)^{\rm T}\in{\bf R}^n$, as above.
Note that
the 
order parameters here are defined by
\bea
q_{wab}\eq\f{1}{N}
\sum_{i=1}^Nw_{ia}w_{ib},\\
q_{sab}\eq\f{1}{N}
\sum_{i=1}^Nv_iw_{ia}w_{ib}.
\eea
In addition,
\bea
\left\langle
f(c,v)
\right\rangle
\eq
\lim_{N\to\infty}
\f{1}{N}
\sum_{i=1}^Nf(c_i,v_i)
\eea
is used.
}

{
In the evaluation 
of 
\siki{eq13},
as the
replica symmetry solution,
\bea
q_{wab}\eq
\left\{
\begin{array}{ll}
\chi_w+q_w&a=b\\
q_w&a\ne b
\end{array}
\right.,\\
q_{sab}\eq
\left\{
\begin{array}{ll}
\chi_s+q_s&a=b\\
q_s&a\ne b
\end{array}
\right.,\\
\tilde{q}_{wab}\eq
\left\{
\begin{array}{ll}
\tilde{\chi}_w-\tilde{q}_w&a=b\\
-\tilde{q}_w&a\ne b
\end{array}
\right.,\\
\tilde{q}_{sab}\eq
\left\{
\begin{array}{ll}
\tilde{\chi}_s-\tilde{q}_s&a=b\\
-\tilde{q}_s&a\ne b
\end{array}
\right.,\\
k_a\eq k,
\eea
are set. From this,
using replica trick 
$\lim_{n\to0}\f{Z^n-1}{n}=\log Z$,
\bea
\phi\eq\lim_{N\to\infty}\f{1}{N}
E_X[\log Z(X)]\nn
\eq\mathop{\rm Extr}_{\theta}
\left\{
\f{1}{2}(\chi_w+q_w)(\tilde{\chi}_w-\tilde{q}_w)+\f{1}{2}q_w\tilde{q}_w
\right.
\nn
&&+\f{1}{2}(\chi_s+q_s)(\tilde{\chi}_s-\tilde{q}_s)+\f{1}{2}q_s\tilde{q}_s-k\nn
&&-\f{\a}{2}\log(1+\b\chi_s)
-\f{\a\b q_s}{2(1+\b\chi_s)}\nn
&&-\f{1}{2}\left\langle
\log(\tilde{\chi}_w+v\tilde{\chi}_s)
\right\rangle+
\f{1}{2}\left\langle
\f{\tilde{q}_w+v\tilde{q}_s}{\tilde{\chi}_w+v\tilde{\chi}_s}
\right\rangle\nn
&&
\left.
+
\f{1}{2}\left\langle
\f{(k-\b\eta c)^2}{\tilde{\chi}_w+v\tilde{\chi}_s}
\right\rangle
\right\}
\eea
is obtained, where the novel set of order parameters $\theta=\left\{\chi_w,q_w,\chi_s,q_s,\tilde{\chi}_w,\tilde{q}_w,
\tilde{\chi}_s,\tilde{q}_s,k
\right\}$ is used.
From these terms in the extremum, the order parameters are\bea
\chi_w\eq\f{\left\langle v^{-1}\right\rangle}{\b(\a-1)},\\
\label{eq20}
\label{eq24}
q_w\eq
\f{1}{\a-1}+\f{\left\langle v^{-2}\right\rangle}
{\left\langle v^{-1}\right\rangle^2}+C(\eta),\\
\chi_s\eq\f{1}{\b(\a-1)},\\
q_s\eq\f{\a}{\a-1}
\left[
\f{1}{\left\langle v^{-1}\right\rangle}
+\f{\eta^2\left\langle v^{-1}\right\rangle V_c}{(\a-1)^2}
\right],\\
\tilde{\chi}_w\eq0,\\
\tilde{q}_w\eq0,\\
\tilde{\chi}_s\eq\b(\a-1),\\
\tilde{q}_s\eq\b^2(\a-1)\left[
\f{1}{\left\langle v^{-1}\right\rangle}
+\f{\eta^2\left\langle v^{-1}\right\rangle V_c}{(\a-1)^2}
\right],\\
k\eq\f{\b(\a-1)}{\left\langle v^{-1}\right\rangle}+\b\eta\f{\left\langle v^{-1}c\right\rangle}{\left\langle v^{-1}\right\rangle},
\eea
where 
\bea
C(\eta)\eq
\f{\eta^2\left\langle v^{-1}\right\rangle^2 V_c}{(\a-1)^3}
+\f{2\eta}{\a-1}
\f{\left\langle v^{-2}\right\rangle}
{\left\langle v^{-1}\right\rangle}
\d_c
\nn
&&
+\f{\eta^2\left\langle v^{-2}\right\rangle }{(\a-1)^2}(V_{cc}+\d_c^2),\\
\label{eq33}
V_c\eq
\f{\left\langle v^{-1}c^2\right\rangle}
{\left\langle v^{-1}\right\rangle}
-\left(
\f{\left\langle v^{-1}c\right\rangle}
{\left\langle v^{-1}\right\rangle}
\right)^2,\\
\d_c\eq 
\f{\left\langle v^{-1}c\right\rangle}
{\left\langle v^{-1}\right\rangle}-
\f{\left\langle v^{-2}c\right\rangle}
{\left\langle v^{-2}\right\rangle},
\\
V_{cc}\eq
\f{\left\langle v^{-2}c^2\right\rangle}
{\left\langle v^{-2}\right\rangle}
-
\left(\f{\left\langle v^{-2}c\right\rangle}
{\left\langle v^{-2}\right\rangle}\right)^2.
\eea
From these and the identity in 
\siki{eq11}, $\ve=-\lim_{\b\to\infty}\pp{\phi}{\b}$,
the minimal investment risk with cost per asset is
\bea
\label{eq32}
\label{eq36}
\ve\eq\lim_{\b\to\infty}
\left\{
\f{\a\chi_s}{2(1+\b\chi_s)}
+\f{\a q_s}{2(1+\b\chi_s)^2}
\right.\nn
&&
\left.
+\left\langle
\f{k-\b\eta c}{\tilde{\chi}_w+v\tilde{\chi}_s}\eta c
\right\rangle
\right\}\nn
\eq
\f{\a-1}{2\left\langle v^{-1}
\right\rangle}
+\eta\f{\left\langle v^{-1}c
\right\rangle}
{\left\langle v^{-1}
\right\rangle}
-\f{\eta^2\left\langle v^{-1}
\right\rangle V_c}{2(\a-1)}.
\eea
Further, 
\siki{eq20} gives the extremal investment concentration $q_w$.}

In the next section, we will discuss numerical experiments conducted in order to 
validate our proposed method. Before then, we should make some comments.
First, a previous work{\cite{PhysRevE.94.062102b}}
has already discussed the portfolio optimization problem 
in the situation that cost when investing is ignored, 
giving the minimal investment risk per asset and 
its investment concentration as follows:\bea
\ve\eq\f{\a-1}{2\left\langle v^{-1}\right\rangle},\\
q_w\eq\f{1}{\a-1}+\f{\left\langle v^{-2}\right\rangle}{\left\langle v^{-1}\right\rangle^2}.
\eea
This corresponds to the case $\eta\to0$ {of our results}.
Next,  for the portfolio optimization problem which minimizes the purchasing cost 
when ignoring investment risk, the purchasing cost is defined as
\bea
{\cal H}'(\vec{w}|X,\vec{c})\eq\sum_{i=1}^Nc_iw_i
\eea
and the minimal cost per asset is
\bea
\ve'\eq\lim_{N\to\infty}\f{1}{N}\mathop{\min}_{\vec{w}\in{\cal W}}{\cal H}'(\vec{w}|X,\vec{c}).
\eea
Then, from the relationship $\ve'=\lim_{\eta\to\infty}\ve/\eta$ and 
using \siki{eq32}, the minimal cost per asset $\ve'$ is obtained {as} $\ve'\to-\infty$.
{This result is supported by the fact that}
there does not exist, for example, a minimum of the function $f(x,y)=2x+3y$ of $x,y$ 
with the two constraint conditions $x+y=1,-\infty<x,y<\infty$. These comments indicate that the findings obtained by the proposed method are 
consistent with the well-known {properties of the optimal solution of the portfolio optimization problem.}

{Lastly,
the minimal expected investment risk with cost $\ve^{\rm OR}$
and its investment concentration $q_w^{\rm OR}$ 
evaluated using the previous analytical procedure (the approach of an annealed disordered system) of operations research
are as follows:
\bea
\label{eq41}
\ve^{\rm OR}\eq
\f{\a}{2\left\langle v^{-1}
\right\rangle}
+\eta\f{\left\langle v^{-1}c
\right\rangle}
{\left\langle v^{-1}
\right\rangle}
-\f{\eta^2\left\langle v^{-1}
\right\rangle V_c}{2\a},\\
q_w^{\rm OR}\eq
\f{\left\langle v^{-2}
\right\rangle
}{\left\langle v^{-1}
\right\rangle^2
}+\f{\eta^2\left\langle v^{-2}
\right\rangle}{\a^2}
(V_{cc}+\d_c^2)
+\f{2\eta}{\a}\f{\left\langle v^{-2}
\right\rangle}{\left\langle v^{-1}
\right\rangle}\d_c.\nn
\label{eq42}
\eea
As an interpretation of this finding, 
since, for example, the function $f(x)=x-\f{b}{x},(x,b>0)$ is 
monotonically increasing in $x$, compared with Eqs. (\ref{eq36})
and (\ref{eq41}), 
\bea
\label{eq43}
\ve&<&\ve^{\rm OR}
\eea
is obtained. That is, in the literature of 
minimization of investment risk with cost,
it has been verified that the minimal investment risk with cost $\ve$
does not correspond to the minimal expected investment risk with cost $\ve^{\rm OR}$; 
{and similarly, }
the investment concentration of the optimal 
$q_w$ {is not equal to} 
the investment concentration of the solution derived in operations research 
$q_w^{\rm OR}$.
}

\section{Numerical Experiments\label{sec5}}
{In this section,
using numerical experiments, 
a verification of the result based on replica analysis
in the preceding section is performed. First,
if the purchasing cost $c_i$ and the variance of return $v_i$ 
do not depend on each other,
then the second term of $\ve$ in \siki{eq36} and 
$V_c$ in \siki{eq33} reduce to
$\f{\left\langle v^{-1}c\right\rangle}
{\left\langle v^{-1}\right\rangle}
={\left\langle c\right\rangle}$
and 
$V_c=\left\langle c^2\right\rangle-\left\langle c\right\rangle^2$.
However, since this model setting is  similar to that of a previous work{\cite{PhysRevE.94.062102b}},
in this paper, we consider the case that $c_i$ and $v_i$ are correlated. 
Here, 
we assume that 
the mean of return $\bar{x}_{i\mu}$, $r_i$, is equal to 
the purchasing cost $c_i$, that is, 
$E[\bar{x}_{i\mu}]=r_i=c_i$.
Moreover, we assume that 
the second moment of return $E[\bar{x}_{i\mu}^2]$
is randomly proportional to 
the square of mean $E[\bar{x}_{i\mu}]$, that is, $E[\bar{x}_{i\mu}^2]=(h_i+1)c_i^2$.
In this setting, the variance of return is 
$V[\bar{x}_{i\mu}]=v_i=h_ic_i^2$. Note that $h_i(>0)$
is the random coefficient and does not depend on $c_i$.
}

{
For the concrete setting of the numerical experiments, 
we assume that 
$c_i,h_i$ are independently distributed {with}
 the bounded Pareto distributions {whose} density functions are denoted by
\bea
\label{eq44}
f_c(c_i)\eq
\left\{
\begin{array}{ll}
\f{(1-b_c)(c_i)^{-b_c}}{(u_c)^{1-b_c}
-(l_c)^{1-b_c}
}&l_c\le c_i\le u_c\\
0&{\rm otherwise}
\end{array}
\right.,\\
f_h(h_i)\eq
\left\{
\begin{array}{ll}
\f{(1-b_h)(h_i)^{-b_h}}{(u_h)^{1-b_h}
-(l_h)^{1-b_h}
}
&l_h\le h_i\le u_h\\
0&{\rm otherwise}
\end{array}
\right.,\qquad
\label{eq45}
\eea
where $u_c,l_c,u_h,l_h$ are the upper and lower bounds of 
$c_i,h_i$, and 
$b_c,b_h(>0)$ are the powers characterizing the
bounded Pareto distributions.
}

{
We here 
do not evaluate analytically 
the inverse matrix $J^{-1}$ in {Eqs. (\ref{eq8}) and (\ref{eq9})} in order to 
assess the optimal portfolio. 
Instead, in the following steps, we derive the optimal portfolio {numerically} by using the steepest descent method
and assess the minimal investment risk with cost $\ve$ and its investment concentration $q_w$.
\begin{description}
\item[Step 1. (Initial setting)]
Assign $c_i$ and $h_i$ randomly according to the density function in \siki{eq44}, $f_c(c_i)$, and 
that in \siki{eq45}, $f_h(h_i)$. 
In particular, random variables $s_c^i,s_h^i$
are independently and identically distributed according to 
the uniform distribution on $[0,1)$, so that 
$c_i=(
s_c^i
(u_c)^{1-b_c}
+(1-s_c^i)
(l_c)^{1-b_c}
)^{\f{1}{1-b_c}}$
 and 
$h_i=(
s_h^i
(u_h)^{1-b_h}
+(1-s_h^i)
(l_h)^{1-b_h}
)^{\f{1}{1-b_h}}$.
\item[Step 2. (Initial setting)]
For asset $i$,
the returns of assets $\bar{x}_{i\mu}$ are 
independently and identically distributed with 
$E[\bar{x}_{i\mu}]=c_i$ and
$V[\bar{x}_{i\mu}]=v_i=h_ic_i^2$.
Moreover, the modified return is
$x_{i\mu}=\bar{x}_{i\mu}-E[\bar{x}_{i\mu}]$. 
Thus, 
the 
return matrix $X=\left\{\f{x_{i\mu}}{\sqrt{N}}\right\}\in{\bf R}^{N\times p}$
is {assigned}.
\item[Step 3. (Initial setting)]
Using the modified return $x_{i\mu}$ in 
Step 2,
\bea
J_{ij}\eq\f{1}{N}
\sum_{\mu=1}^p
x_{i\mu}x_{j\mu}.
\eea
\item[Step 4. (Initial setting)]
Set the initial portfolio 
$\vec{w}$ and 
Lagrange coefficient $k$ as 
$\vec{w}_0=\vec{e}=(1,1,\cdots,1)^{\rm T}\in{\bf R}^N$
and $k_0=1$ and the initial of cost tolerance $\eta$ 
as $\eta_{\min}$.
\item[Step 5. (Optimization)]
Using the portfolio at iteration step $t$, 
$\vec{w}_t=(w_{1,t},w_{2,t},\cdots,w_{N,t})^{\rm T}\in{\bf R}^N$,
and Lagrange coefficient $k_t$, 
update $\vec{w}_{t+1}$ (the portfolio at iteration step $t+1$) 
 and the Lagrange coefficient 
$k_{t+1}$ (using the steepest descent method for $L(\vec{w},k)$ in \siki{eq8-1}) as follows:
\bea
\vec{w}_{t+1}\eq \vec{w}_t-\g_w
\left(
\pp{L(\vec{w},k)}{\vec{w}}\right)_{\vec{w}=\vec{w}_t,k=k_t},\qquad\\
k_{t+1}\eq k_t+\g_k
\left(
\pp{L(\vec{w},k)}{k}\right)_{\vec{w}=\vec{w}_t,k=k_t},
\eea
where $\g_w,\g_k(>0)$ are the learning rates of the steepest descent method.
\item[Step 6. (Optimization)]
Compute the difference between $\vec{w}_t,k_t$ and 
$\vec{w}_{t+1},k_{t+1}$,
\bea
\Delta\eq
\sum_{i=1}^N
\left|w_{i,t}-w_{i,t+1}\right|
+|k_t-k_{t+1}|.
\eea
\item[Step 7. (Optimization)]
If $\Delta>\d$, 
then update $t\leftarrow t+1$ and go back to Step 5. If $\Delta<\d$,
then, regarding $\vec{w}_{t+1}$ and 
$k_{t+1}$ as the approximations of the optimal 
portfolio $\vec{w}^*=\arg\mathop{\min}_{\vec{w}\in{\cal W}}{\cal H}(\vec{w}|X,\vec{c})$
and Lagrange coefficient $k^*$, evaluate 
the minimal investment risk with cost per asset 
$\ve(\eta,X)$ and its investment concentration 
$q_w(\eta,X)$,
and go to Step 8.
\item[Step 8. (Optimization)]
If $\eta+d_\eta<\eta_{\max}$, then update
$\eta\leftarrow\eta+d_\eta$ and go back to Step 5.
If $\eta+d_\eta>\eta_{\max}$, then stop the steepest descent algorithm.
\end{description}
}

{
Note that 
we do not use either the replica symmetry ansatz and a calculation of an inverse matrix
in this algorithm. Moreover, using this steepest descent method algorithm $M$ times, 
with respect to the return matrix assigned in the initial setting of the $m(=1,2,\cdots,M)$th trial, 
$X^m=\left\{\f{x_{i\mu}^m}{\sqrt{N}}\right\}\in{\bf R}^{N\times p}$, which is used to assess 
the minimal investment risk with cost $\ve(\eta,X^m)$ and its investment concentration 
$q_w(\eta,X^m)$ and
the sample averages of the minimal investment risk with cost per asset and
the investment concentration of the optimal portfolio are 
\bea
\ve(\eta)\eq\f{1}{M}\sum_{m=1}^M\ve(\eta,X^m),\\
q_w(\eta)\eq\f{1}{M}\sum_{m=1}^Mq_w(\eta,X^m),
\eea
where $\ve(\eta,X^m)$ and $q_w(\eta,X^m)$ are the results of $m$th trial.
}

\begin{figure}[t] 
\begin{center}
\includegraphics[width=1.0\hsize]{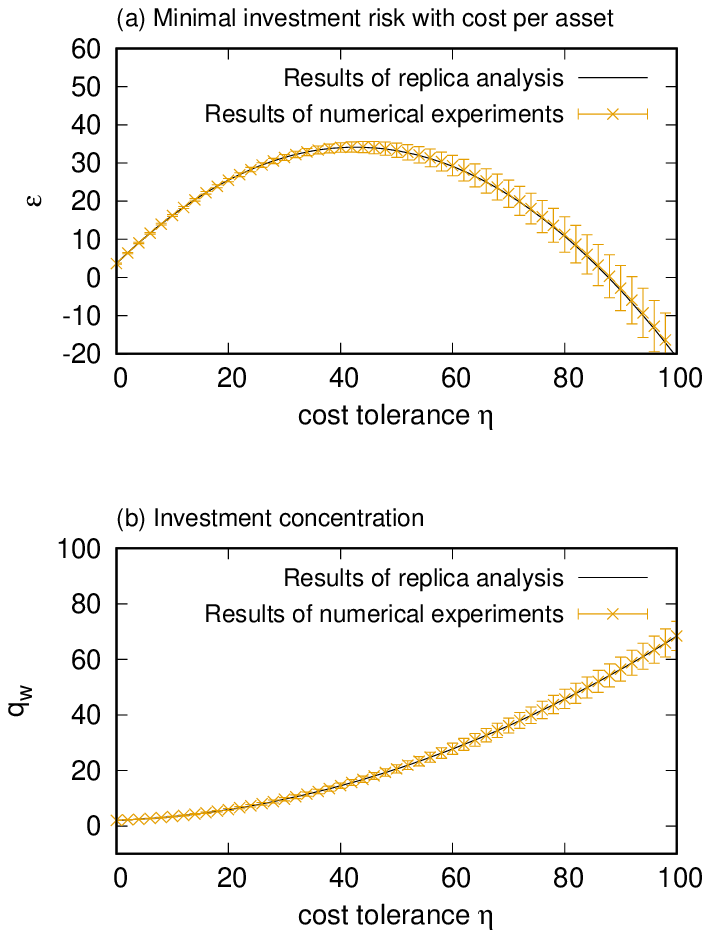}
\caption{
\label{Fig1}
{
Results of the replica analysis and the numerical experiments ($\a=p/N=3$).
The horizontal axis indicates the cost tolerance $\eta$, and the vertical axes show 
(a) the minimal investment risk with cost per asset $\ve$, and (b) the investment 
 concentration $q_w$. 
The black {solid} lines indicate the results of 
the replica analysis for (a) 
 \siki{eq36} and (b) \siki{eq20}. The orange crosses with 
 error bars indicate the results of the numerical simulations. 
}
}
\end{center}
\end{figure}

{For the numerical simulations, 
$N=1000,p=3000, (\a=p/N=3)$, and the parameters of the bounded 
Pareto distribution are $(b_c,u_c,l_c)
=
(b_h,u_h,l_h)=(2,4,1)
$.
Further, $(\eta_{\min},\eta_{\max},d_\eta)=(0,100,2)$ defines the range of cost tolerance $\eta$ and its increment,
the learning rates of the steepest descent method are $\g_w=\g_k=10^{-3}$,
and the constant of 
the stopping condition is $\d=10^{-6}$. 
Finally, the total number of trials is $M=100$. The
numerical results estimated by this steepest 
descent method with these numerical settings (orange crosses with 
error bars) 
and those based on replica analysis (black {solid} lines)
are shown in Fig. \ref{Fig1}.
As shown, the results derived by using replica analysis 
and the numerical results 
are consistent with each other, which verifies 
{the validity of our proposed method based on replica analysis.} 
In addition, 
{from Eqs. (\ref{eq43}), (\ref{eq24}), and (\ref{eq42}),}
the analytical approach developed in operations research in previous works 
is difficult to use to examine
the minimization problem of
the investment risk with cost under a budget constraint, that is,
it is disclosed that the analytical approach developed in operations research
cannot examine 
the properties of the minimal investment risk with cost and 
the investment concentration {of the optimal portfolio.}
}

\section{Conclusion\label{sec6}}
{In this
work, we have investigated using replica analysis 
the 
minimization problem of the investment risk with cost which is 
defined by two types of loss in investment, the risk and cost.
Concretely, based on mathematical similarity,
we regarded the investment risk with cost 
as the Hamiltonian of this investment system, and further,
since 
this system is mathematically analogous to
the Hamiltonians of the Hopfield model and the SK model,
we recognized that we could analyze the portfolio optimization problem using replica analysis.
Similar to in previous {works\cite{Ciliberti2007,doi:10.1080/14697680701422089,PAFKA2003487,KONDOR20071545,
doi:10.1080/1351847X.2011.601661,10.1371/journal.pone.0134968,10.1371/journal.pone.0133846,1742-5468-2017-2-023301,SHINZATO2018986,
doi:10.7566/JPSJ.86.124804,PhysRevE.94.052307,1742-5468-2016-12-123404,2017arXiv170302777S,PhysRevE.94.062102b,
2016arXiv160506840S,doi:10.7566/JPSJ.86.063802,110008689817,1742-5468-2017-12-123402},
}
we were able to 
examine 
the minimal investment risk with cost and the investment concentration of 
the optimal portfolio minimizing 
the investment risk with cost thoroughly  based on the replica symmetry ansatz.
In addition, 
we showed 
that the minimal investment risk with cost
and its investment concentration which are evaluated by the 
approach of a quenched disordered system
are in no way consistent with
the minimal expected investment risk with cost 
and the investment concentration minimizing the expected investment risk with cost
which are evaluated by
the approach developed in operations research (that is, the approach of an annealed disordered system).
Using the results of numerical simulations,
we verified the validity of our proposed method based on replica analysis. Namely,
we showed that the properties of the minimal investment risk with cost 
and its investment concentration, 
which are not easily analyzed by the analytical approach 
developed in operations research, 
are revealed by the quenched disordered approach.
}

{In this paper, we assumed that, 
with respect to the cost per unit portfolio, 
purchasing cost is equal to selling cost;
however, as
future research, we also need to considered the case that 
purchasing cost $c_i$ (the cost on $w_i>0$) and 
selling cost $c_i'$ (the cost on $w_i<0$) are distinct.
For this purpose, as a generalization, we need to consider the portfolio optimization problem for the case that 
the cost needs to be represents as a piecewise linear 
or nonlinear function;  for example, we can change $\sum_{i=1}^Nc_iw_i$ in \siki{eq3} to
$\sum_{i=1}^N(c_i\max(w_i,0)-c_i'\max(-w_i,0))$. 
Moreover, 
in order to construct a macroscopic relation of the diversification investment theory,
we need to derive a relation 
between the macroscopic variables like the Pythagorean theorem of the Sharpe ratio and 
the relation of loss opportunity
{\cite{2017arXiv170302777S,PhysRevE.94.062102b,
2016arXiv160506840S}}. Further,
so as to examine the properties of the utility function 
of the optimal portfolio, we need to investigate 
several performance indicators rather than merging risk and cost
(see appendix \ref{app-b}).
}

\section*{Acknowledgements}
{
The author is grateful for detailed discussions with
K. Kobayashi, H. Yamamoto, and D. Tada. 
This work was supported in part by
Grants-in-Aid Nos. 15K20999, 17K01260, and 17K01249; Research Project of the Institute of Economic Research Foundation at Kyoto University; and Research Project
No. 4 of the Kampo Foundation.
}

\appendix
\section{Moments\label{app-a}}
{In this appendix,
using replica analysis, 
we will {calculate} the six moments in the argument of Lagrange multiplier's method, 
$\f{1}{N}\vec{e}^{\rm T}J^{-1}\vec{e}$, 
$\f{1}{N}\vec{e}^{\rm T}J^{-1}\vec{c}$, 
$\f{1}{N}\vec{c}^{\rm T}J^{-1}\vec{c}$, 
$\f{1}{N}\vec{e}^{\rm T}J^{-2}\vec{e}$, 
$\f{1}{N}\vec{e}^{\rm T}J^{-2}\vec{c}$, and
$\f{1}{N}\vec{c}^{\rm T}J^{-2}\vec{c}$. First, the following partition $Z(y,X)$ is applied:
\bea
Z(y,X)\eq\f{1}{(2\pi)^{\f{N}{2}}}
\area d\vec{w}e^{-\f{1}{2}\vec{w}^{\rm T}(J-yI_N)\vec{w}+k\vec{w}^{\rm T}\vec{e}+
\theta\vec{w}^{\rm T}\vec{c}},\nn
\eea
where $J=XX^{\rm T}$. Further, we analyze
\bea
&&\log Z(y,X)\nn
\eq-\f{1}{2}\log\det|J-yI_N|
+\f{k^2}{2}\vec{e}^{\rm T}(J-yI_N)^{-1}\vec{e}
\nn
&&+\f{\theta^2}{2}\vec{c}^{\rm T}(J-yI_N)^{-1}\vec{c}+k\theta\vec{e}^{\rm T}(J-yI_N)^{-1}\vec{c}.\nn
\eea
For this purpose, we define\bea
\phi(y)\eq\lim_{N\to\infty}\f{1}{N}\log Z(y,X).
\eea
{Thus,} $\phi(0)$ and $\phi'(0)$ are given by
\bea
\phi(0)\eq-\f{1}{2}\lim_{N\to\infty}\f{1}{N}\log\det|J|+\f{k^2}{2}\lim_{N\to\infty}\f{\vec{e}^{\rm T}J^{-1}\vec{e}}{N}\nn
&&+\f{\theta^2}{2}\lim_{N\to\infty}\f{\vec{c}^{\rm T}J^{-1}\vec{c}}{N}
+k\theta\lim_{N\to\infty}\f{\vec{e}^{\rm T}J^{-1}\vec{c}}{N},\\
\phi'(0)\eq\f{1}{2}\lim_{N\to\infty}\f{1}{N}{\rm Tr}J^{-1}+\f{k^2}{2}\lim_{N\to\infty}\f{\vec{e}^{\rm T}J^{-2}\vec{e}}{N}\nn
&&+\f{\theta^2}{2}\lim_{N\to\infty}\f{\vec{c}^{\rm T}J^{-2}\vec{c}}{N}
+k\theta\lim_{N\to\infty}\f{\vec{e}^{\rm T}J^{-2}\vec{c}}{N}.\qquad\ 
\eea
The second derivatives of $\phi(0)$ and $\phi'(0)$ with respect to 
$k,\theta$ allow the six moments to be analyzed exactly.
Moreover, in a similar way to that used in a previous work{\cite{PhysRevE.94.052307}},
since the logarithm of the partition function maintains the property of self-averaging,
using replica analysis and the replica symmetric ansatz, 
\bea
\phi(y)\eq
\lim_{N\to\infty}
\f{1}{N}E_X[\log Z(y,X)]\nn
\eq
\mathop{\rm Extr}_{\chi_s,q_s,\tilde{\chi}_s,\tilde{q}_s}
\left\{
-\f{\a}{2}\log(1+\chi_s)
-\f{\a q_s}{2(1+\chi_s)}
\right.
\nn
&&+\f{1}{2}(\chi_s+q_s)(\tilde{\chi}_s-\tilde{q}_s)+\f{1}{2}q_s\tilde{q}_s
\nn
&&
-\f{1}{2}
\left\langle
\log (v\tilde{\chi}_s-y)
\right\rangle+
\f{1}{2}
\left\langle
\f{v\tilde{q}_s}{v\tilde{\chi}_s-y}
\right\rangle
\nn
&&
\left.
+\f{1}{2}
\left\langle
\f{(k+c\theta)^2}{v\tilde{\chi}_s-y}
\right\rangle
\right\},
\label{eq-a6}
\eea
is assessed as follows.
From the extremum of the order parameters when $y=0$,
we analytically derive $\chi_s=\f{1}{\a-1},\tilde{\chi}_s=\a-1,q_s=\f{\a}{(\a-1)^3}\left\langle
\f{(k+c\theta)^2}{v}
\right\rangle$, and $\tilde{q}_s=\f{1}{\a-1}\left\langle
\f{(k+c\theta)^2}{v}
\right\rangle$. Substituting these into \siki{eq-a6},
\bea
\phi(0)\eq-\f{\a}{2}\log\f{\a}{\a-1}-\f{1}{2}\log(\a-1)+\f{1}{2}-\f{1}{2}
\left\langle
\log v
\right\rangle\nn
&&+\f{1}{\a-1}\left\langle
\f{(k+c\theta)^2}{v}
\right\rangle,
\eea
and
\bea
\phi'(0)\eq
\f{\left\langle v^{-1}\right\rangle}{2\tilde{\chi}_s}
+
\f{\left\langle v^{-1}\right\rangle}{2}\f{\tilde{q}_s}{\tilde{\chi}_s^2}
+\f{1}{2\tilde{\chi}_s^2}
\left\langle 
\f{(k+c\theta)^2}{v^2}
\right\rangle\nn
\eq\f{\left\langle v^{-1}\right\rangle}{2(\a-1)}+
\f{\left\langle v^{-1}\right\rangle}{2(\a-1)^3}
\left\langle 
\f{(k+c\theta)^2}{v}
\right\rangle\nn
&&
+
\f{1}{2(\a-1)^2}
\left\langle 
\f{(k+c\theta)^2}{v^2}
\right\rangle,
\eea
where $\tilde{\chi}_s=\a-1,\tilde{q}_s=
\f{1}{\a-1}\left\langle
\f{(k+c\theta)^2}{v}
\right\rangle
$ have already been substituted. From this,
we can evaluate the second derivatives of $\phi(0)$ and $\phi'(0)$ with respect to 
$k,\theta$ analytically as
\bea
\label{eq-a9}
\lim_{N\to\infty}
\f{1}{N}\vec{e}^{\rm T}J^{-1}\vec{e}
\eq\pp{^2\phi(0)}{k^2}\nn
\eq\f{\left\langle
v^{-1}
\right\rangle
}{\a-1},\\
\lim_{N\to\infty}
\f{1}{N}\vec{e}^{\rm T}J^{-1}\vec{c}
\eq\pp{^2\phi(0)}{k\p \theta}\nn
\eq\f{\left\langle
v^{-1}c
\right\rangle
}{\a-1},\\
\label{eq-a11}
\lim_{N\to\infty}
\f{1}{N}\vec{c}^{\rm T}J^{-1}\vec{c}
\eq\pp{^2\phi(0)}{\theta^2}\nn
\eq\f{\left\langle
v^{-1}c^2
\right\rangle
}{\a-1}\\
\lim_{N\to\infty}
\f{1}{N}\vec{e}^{\rm T}J^{-2}\vec{e}
\eq\pp{^2\phi'(0)}{k^2}\nn
\eq\f{\left\langle
v^{-1}
\right\rangle^2
}{(\a-1)^3
}+
\f{\left\langle
v^{-2}
\right\rangle
}{(\a-1)^2
}
,\\
\lim_{N\to\infty}
\f{1}{N}\vec{e}^{\rm T}J^{-2}\vec{c}
\eq\pp{^2\phi'(0)}{k\p \theta}\nn
\eq\f{\left\langle
v^{-1}
\right\rangle
\left\langle
v^{-1}c
\right\rangle
}{(\a-1)^3
}+
\f{\left\langle
v^{-2}c
\right\rangle
}{(\a-1)^2
}
,\nn
\\
\lim_{N\to\infty}
\f{1}{N}\vec{c}^{\rm T}J^{-2}\vec{c}
\eq\pp{^2\phi'(0)}{\theta^2}\nn
\eq
\f{\left\langle
v^{-1}
\right\rangle
\left\langle
v^{-1}c^2
\right\rangle
}{(\a-1)^3
}+
\f{\left\langle
v^{-2}c^2
\right\rangle
}{(\a-1)^2
}.\nn
\label{eq-a14}
\eea
Next,
using the result of \siki{eq8} in the case of a finite number of assets $N$, 
in the thermodynamical limit of $N$, these should 
maintain the self-averaging property, 
so we substitute the results in Eqs. (\ref{eq-a9}) to (\ref{eq-a11}) into
(\ref{eq8}),
\bea
\ve\eq
\f{1}{2}\f{\left(1+\eta\f{\left\langle
v^{-1}c
\right\rangle
}{\a-1}\right)^2}
{\f{\left\langle
v^{-1}
\right\rangle
}{\a-1}}-\f{\eta^2}{2}
\f{\left\langle
v^{-1}c^2
\right\rangle
}{\a-1}\nn
\eq
\f{\a-1}{2\left\langle
v^{-1}
\right\rangle
}+\eta
\f{\left\langle
v^{-1}c
\right\rangle
}
{\left\langle
v^{-1}
\right\rangle
}
-\f{\eta^2\left\langle
v^{-1}
\right\rangle}
{2(\a-1)}
V_c,
\eea
which is consistent with the result based on replica analysis in
\siki{eq32}. Similarly, if the results 
from Eqs. (\ref{eq-a9}) to (\ref{eq-a14}) are substituted into 
\siki{eq9}, then it is also verified that this result corresponds to that based on replica analysis in 
\siki{eq20}.
}
\section{Investment risk with return and cost\label{app-b}}
Since the model handled in this paper 
is mathematically analogous to both the Hopfield model and the SK model, we have focused on the minimization problem of the investment risk with cost.
Here, however, let us consider the minimization problem of 
the investment risk with return, which has been widely investigated 
in operations research. First, the expected return of the portfolio $\vec{w}$ is
defined as follows:
\bea
Return\eq
\sum_{i=1}^Nw_ir_i,
\eea
where $r_i$ is the mean of return of asset $i$, that is, $E[\bar{x}_{i\mu}]=r_i$. In
this setting, the investment risk with return is 
\bea
{\cal H}(\vec{w}|X,\vec{r})
\eq
\f{1}{2}\vec{w}^{\rm T}J\vec{w}-g\vec{r}^{\rm T}\vec{w},
\eea
where $g(>0)$ is the mixing degree of return. From this,
when 
$\eta c_i$ in the main manuscript is replaced by $-gr_i$, 
the minimal investment risk with return 
per asset is $\ve=\lim_{N\to\infty}\f{1}{N}
\mathop{\min}_{\vec{w}\in{\cal W}}{\cal H}(\vec{w}|X,\vec{r})
$ based on 
\siki{eq36}. Then, 
\bea
\ve\eq
\f{\a-1}{2\left\langle v^{-1}
\right\rangle}
-g\f{\left\langle v^{-1}r
\right\rangle}
{\left\langle v^{-1}
\right\rangle}
-\f{g^2\left\langle v^{-1}
\right\rangle V_r}{2(\a-1)},
\eea
where 
\bea
V_r\eq
\f{\left\langle v^{-1}r^2\right\rangle}
{\left\langle v^{-1}\right\rangle}
-\left(
\f{\left\langle v^{-1}r\right\rangle}
{\left\langle v^{-1}\right\rangle}
\right)^2.
\eea
In addition, we can also consider the minimization problem of 
the investment risk 
with
both return and cost added, that is,
the investment risk 
with return and cost, as follows:
\bea
{\cal H}(\vec{w}|X,\vec{r},\vec{c})
\eq
\f{1}{2}\vec{w}^{\rm T}
J\vec{w}
-g\vec{r}^{\rm T}\vec{w}
+\eta\vec{c}^{\rm T}\vec{w}
.
\eea
Then the minimal investment risk with return and cost per asset 
$\ve=
\lim_{N\to\infty}\f{1}{N}
\mathop{\min}_{\vec{w}\in{\cal W}}{\cal H}(\vec{w}|X,\vec{r},\vec{c})
$ can be calculated as
\bea
\ve
\eq\f{\a-1}{2\left\langle v^{-1}
\right\rangle}
-g\f{\left\langle v^{-1}r
\right\rangle}
{\left\langle v^{-1}
\right\rangle}
+\eta\f{\left\langle v^{-1}c
\right\rangle}
{\left\langle v^{-1}
\right\rangle}
-\f{\left\langle v^{-1}
\right\rangle V}{2(\a-1)}
,\nn
\\
V\eq
\f{\left\langle v^{-1}(\eta c-gr)^2\right\rangle}
{\left\langle v^{-1}\right\rangle}
-\left(
\eta\f{\left\langle v^{-1}c
\right\rangle}
{\left\langle v^{-1}
\right\rangle}
-g\f{\left\langle v^{-1}r
\right\rangle}
{\left\langle v^{-1}
\right\rangle}
\right)^2,\nn
\eea
where 
$\eta c_i$ in \siki{eq4} is replaced by 
$-gr_i+\eta c_i$. This shows that 
we can analyze a utility function which comprises risk, return, and cost. Note that the utility function
depends on the preferences of each investor; that is, 
the utility function is a subjective criterion based on each individual's needs
of what the important factors are for the investor to decide to invest.
As individual terms in the utility function, 
it is well known that the utility function 
may include risk, return, and cost (that is, 
the mixing degree of return $g$ and cost tolerance $\eta$
differ between investors). As mentioned in the main manuscript, investment 
theory should be deepened in order to meet the needs of each investor and 
an optimal investment strategy should be proposed for the rational investor.

\bibliographystyle{jpsj}
\bibliography{sample20180118}

\end{document}